\documentclass[twocolumn,aps,showpacs,prl,groupedaddress]{revtex4}
\usepackage{epsfig,amssymb,amsmath}
\usepackage[hypertex,linkcolor=red]{hyperref}

\usepackage{color}

\def\comment#1{}
\def\labell#1{\label{#1}}
\def\sectionp#1{{\par\em #1:--- }}

\def\togli#1{}

\def\>{\rangle}
\def\<{\langle}
\begin{document}

\title{Complementarity and correlations} \author{Lorenzo Maccone$^1$, Dagmar Bru\ss$^2$,
  Chiara Macchiavello$^1$}
\affiliation{\vbox{$^1$Dip.~Fisica and INFN
    Sez.~Pavia, 
    University~of Pavia, via Bassi 6, I-27100 Pavia,
    Italy}\\\vbox{$^2$Institut f\"ur Theoretische Physik III, Heinrich-Heine-Universit\"at D\"usseldorf,
40225 D\"usseldorf, Germany}
}
\begin{abstract}
  We provide an interpretation of entanglement based on classical
  correlations between measurement outcomes of complementary
  properties: states that have correlations beyond a certain threshold
  are entangled. The reverse is not true, however. We also show that,
  surprisingly, all separable nonclassical states exhibit smaller
  correlations for complementary observables than some strictly classical
  states. We use mutual information as a measure of classical
  correlations, but we conjecture that the first result holds also for
  other measures (e.g.~the Pearson correlation coefficient or the sum
  of conditional probabilities).
\end{abstract}
\togli{
We investigate the relation between complementary properties
and correlations of composite quantum systems. We introduce
three measures of correlations which are based on local
measurements in complementary bases. These measures are linked to
the mutual information, the Pearson correlation coefficient
and the sum of conditional probabilities, respectively.
We show that states which have complementary correlations
beyond a certain threshold must be entangled. The reverse
is not true, however. We also show that, surprisingly,
states with non-zero quantum correlations may have less
correlations on complementary observables than classically
correlated states.
}
\togli{  We show that states that have more correlations among complementary
  observables must be entangled. The reverse is false: general
  entangled states do not have more correlations on complementary
  observables than separable ones. We either prove or conjecture that
  this is true for different measures of correlation: the mutual
  information, the Pearson correlation coefficient, and the sum of
  conditional probabilities. We also show that, by this measure,
  surprisingly states with nonzero discord may have less correlation
  on complementary observables than classically correlated states.}

%

\maketitle

Two properties of a quantum state are called complementary if they are
such that, if one knows the value of one property, all possible values
of the other property are equiprobable. More rigorously, let $|a_i\>$
represent the eigenstates corresponding to possible values of a
nondegenerate property $A=\sum_if(a_i)|a_i\>\<a_i|$, and $|c_i\>$ the
eigenstates of a nondegenerate property $C=\sum_jg(c_j)|c_j\>\<c_j|$
(with $f$ and $g$ arbitrary bijective functions). Then $A$ and $B$ are
complementary properties if for all $i,j$ we have
$|\<a_i|c_j\>|^2=1/d$, $d$ being the Hilbert space dimension. Clearly
complementary properties with this definition identify two mutually
unbiased bases (MUBs) \cite{mubs}. Here we study what classical
correlations in the
measurements of these complementary properties tell us about the
quantum correlations of the state of the system.

Typically one discusses entanglement \cite{review} 
in terms of non-locality, Bell inequality violations, monotones over
LOCC, etc. For example, previous literature on entanglement focused on
time-reversal (for the PPT criterion \cite{pereshor,wern}), local
uncertainty relations \cite{olga,mahler,gunnar,guehne06,zhang07},
entropic uncertainty relations \cite{vittorio,gl04,yichen,coles},
entanglement witnesses \cite{ter,cir,geza,spengler}, concurrence
\cite{bill}, the cross-norm criterion \cite{rudolph} and the
covariance matrix criterion \cite{guhne,vicen,eisert,eisert1,zhang}
(the latter encompassing many of the former).  In contrast to these
studies, we focus specifically on classical correlations for
complementary properties. \togli{All these studies are inequivalent to
  the analysis proposed here, since our method is based on
  correlations among {\em two} complementary properties (and can be
  easily extended to more).  } Classical correlations are typically
quantified in terms of the mutual information, which is the main
quantity considered here. We will also discuss the case of alternative
measures such as the Pearson correlations and the sum of conditional
probabilities.  In \cite{wu} related approaches using specific
measures of correlations (different from the ones used here) were
proposed.

The outline of the paper follows. We start by describing the general
scenario we employ for correlation evaluation. We then introduce
different measures of correlations and state our results and our
conjectures regarding entanglement and quantum correlations. We
provide some examples of applications. The details of the proofs of
our results are reported in the supplemental material.

\begin{figure}[hbt]
\begin{center}
\epsfxsize=.8\hsize\leavevmode\epsffile{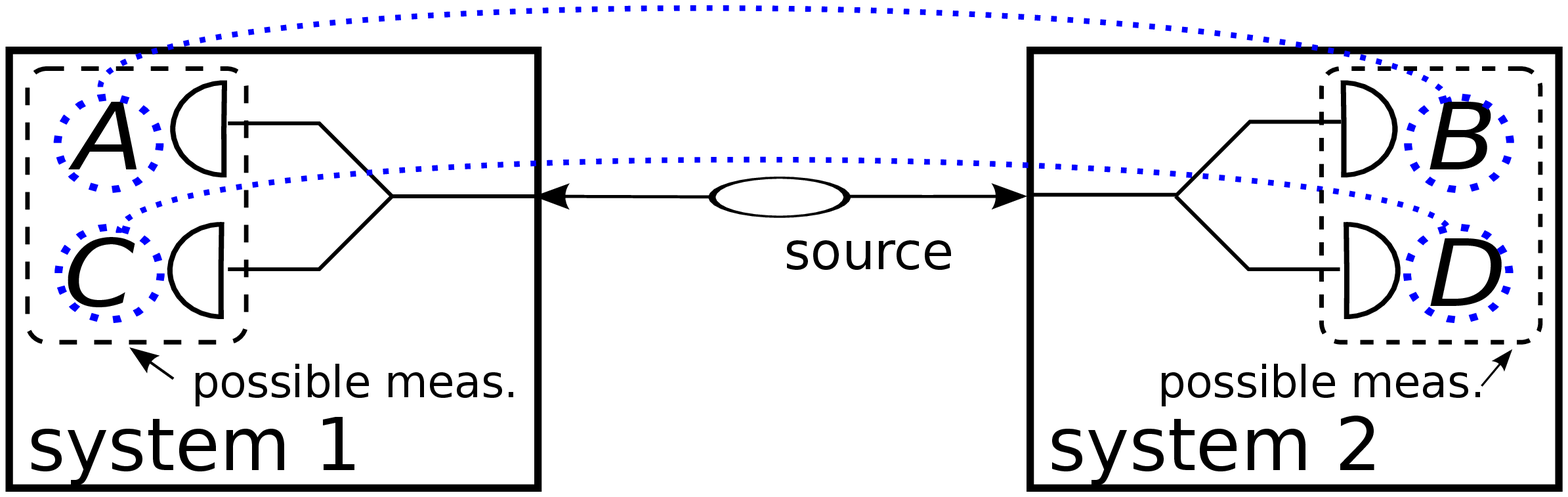}
\end{center}
\vspace{-.5cm}
\caption{Complementary correlation measurements. Each of two systems
  is subject to the measurement of one of two observables: either $A$
  or $C$ on system 1 and either $B$ or $D$ on system 2. Correlations
  are evaluated between the results of $A$ and $B$ and between $C$ and
  $D$ (dashed lines). $A$ and $C$ are complementary on the first
  system, $B$ and $D$ on the second.}
\labell{f:illust}\end{figure}
\sectionp{Complementary correlations} Consider two systems of finite
dimension $d$(\footnote{Many results can be immediately extended to
  the case where the two systems have different dimensions.}) and two
observables $A\otimes B$ and $C\otimes D$ (Fig.~\ref{f:illust}) where
$A$ and $C$ are complementary on the first system (namely
$|\<a_i|c_j\>|=1/\sqrt{d}$ for all eigenstates of $A$ and $C$) and $B$
and $D$ on the second. For example, take the computational basis of
the two systems as the eigenstates of $A$ and $B$, and the Fourier
basis as the ones of $C$ and $D$. We can quantify the correlations
between the results of measurements of $A$ and $B$ with some
correlation measure ${\cal X}_{AB}$ and the correlations between $C$
and $D$ with ${\cal X}_{CD}$. As ${\cal X}$ below 
we will define and investigate three possibilities: the
mutual information ${\cal X}_{XY}=I_{XY}$, the sum of conditional
probabilities ${\cal X}_{XY}={\cal S}_{XY}$, and the Pearson
correlation coefficient ${\cal X}_{XY}={\cal C}_{XY}$. A measure of
the overall correlation of the initial state, which we name the ``complementary
correlations'', can then be given as the sum of the absolute value of
the two measures $|{\cal X}_{AB}|+|{\cal X}_{CD}|$ or as the product $|{\cal
  X}_{AB}{\cal X}_{CD}|$. The latter is typically a weaker measure
than the former, since an upper bound for the sum implies an upper
bound for the product.  Indeed, $(|{\cal X}_{AB}|^{1/2}-|{\cal
  X}_{CD}|^{1/2})^2\geqslant 0$ implies $2\sqrt{|{\cal X}_{AB}{\cal
    X}_{CD}|}\leqslant|{\cal X}_{AB}|+|{\cal X}_{CD}|$. Thus we will
mainly consider the sum of correlations for complementary observables
$|{\cal X}_{AB}|+|{\cal X}_{CD}|$ as a way to evaluate the
complementary correlations. 

\sectionp{Mutual Information} We start considering the mutual
information: $I_{AB}\equiv H(A)-H(A|B)$, where $H(A)$ is the Shannon
entropy of the probabilities of the measurement outcomes of the first
system and $H(A|B)$ is the conditional entropy of the outcomes of the
first conditioned on the second. The complementary correlations are
then $I_{AB}+I_{CD}$.

The relation of this quantity to the entanglement and the discord of
the state of the system is illustrated by the following results:
(i)~The state of a bipartite composite quantum system is maximally
entangled if and only if there exist two complementary measurement
bases where $I_{AB}+I_{CD}=2\log_2d$; (ii)~If \begin{eqnarray}
I_{AB}+I_{CD}>\log_2
d,
\labell{a}\;
\end{eqnarray}
the state of the bipartite system is entangled (see also
\cite{james}); (iii)~The separable states that satisfy this inequality
with equality (i.e.~$I_{AB}+I_{CD}=\log_2 d$), are the
classically-correlated (CC) zero-discord states of the form
\begin{eqnarray}
\rho_{cc}=\sum_i|a_i\>\<a_i|\otimes|b_i\>\<b_i|/d
\labell{rhoc}\;
\end{eqnarray}
with $|a_i\>$ and $|b_i\>$ eigenstates of $A$ and $B$ (or the
analogous state with a uniform convex combination of eigenstates of
$C$ and $D$). Some examples of $I_{AB}+I_{CD}$ for various families of
states are plotted in Fig.~\ref{f:example}a, where we emphasize the
threshold $\log_2d$ above which all states are entangled.

The first result follows from the fact that each term in the sum is
upper bounded by $\log_2d$ by definition. The maximum value for the
sum is then $2\log_2d$ and is achievable if and only if there is
maximal correlation both between $A$ and $B$, and between $C$ and $D$.
Simple properties of the conditional probabilities (see supplemental
material) imply that this can happen for a suitable choice of
observables if and only if the state is maximally entangled. The
second result is a consequence of the concavity of the entropy and of
Maassen and Uffink's entropic uncertainty relation \cite{mu} (see
supplemental material for the details). It gives a sufficient
condition for entanglement that can be used for entanglement
detection.  The third result is surprising: one might expect that the
separable states at the boundary with the entangled region are highly
quantum correlated, whereas we find that they only have classical
correlations (CC) and no discord. This means that quantum correlated
states without entanglement do not have higher correlations for
complementary properties than CC states.  This result is peculiar for
the mutual information as a figure of merit, it is no longer true for
the Pearson correlation (where a family of QQ states sits on the
border, as shown in Fig.~\ref{f:example}b). It can be proved by
analyzing the conditions for equality of the concavity of the entropy
and of Maassen and Uffink's inequality (see supplemental material).

\begin{figure}[hbt]
\begin{center}
\epsfxsize=1.\hsize\leavevmode\epsffile{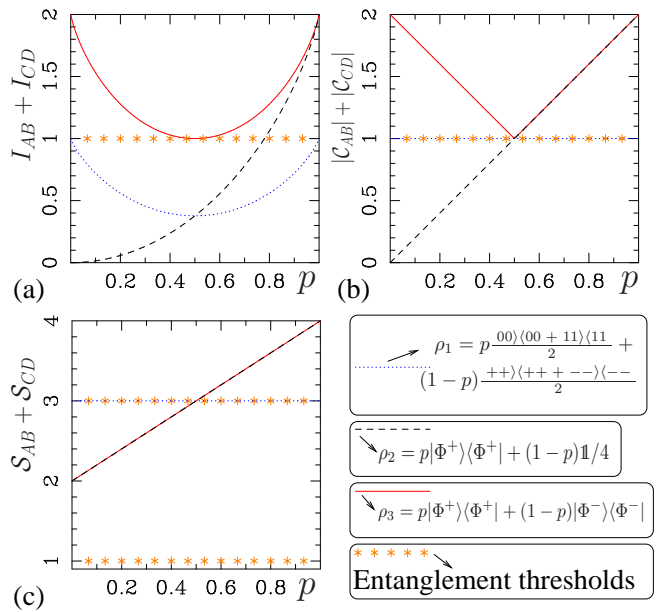}
\end{center}
\vspace{-.5cm}
\caption{Examples of complementary correlations for different measures
  of correlation and different families of states. (a)~Correlation
  $I_{AB}+I_{CD}$ plotted as a function of the parameter $p$ for the
  families of $p$-dependent two-qubit states indicated in the lower
  right panel.  The dotted-line states are always separable and are
  nonzero discord QQ states for $p\neq 0,1$, the dashed-line states
  (Werner states) are entangled for $p>1/3$, whereas the solid-line
  states are entangled for $p\neq1/2$ . Above the threshold 1 (stars)
  the states are certainly entangled. (b)~Same as previous for $|{\cal
    C}_{AB}|+|{\cal C}_{CD}|$, note that the QQ state (dotted line) is
  on the conjectured threshold 1 (stars) for this measure of
  correlation.  (c)~Same as previous for ${\cal S}_{AB}+{\cal
    S}_{CD}$.  Here there are two entanglement boundaries: the states
  that have sum larger than $3$ or smaller than $1$ are conjectured to
  be entangled.  Again, the dotted-state coincides with one of the
  conjectured boundaries. The dashed-line and the solid line are
  superimposed.  Here $|\pm\>\equiv(|0\>\pm|1\>)/\sqrt{2}$ and
  $|\Phi^{\pm}\>\equiv(|00\>\pm|11\>)/\sqrt{2}$. }
\labell{f:example}\end{figure}

\sectionp{Pearson correlation} The second measure of correlation we
consider is the Pearson correlation coefficient ${\cal C}_{AB}$,
defined as
\begin{eqnarray}
{\cal  C}_{AB}\equiv\frac{\<AB\>-\<A\>\<B\>}{\sigma_A\sigma_B}\;, 
\labell{C_AB}
\end{eqnarray}
where, as before, $A$ and $B$ denote observables relative to the two
systems, $\<X\>=$Tr$[X\rho]$ is the expectation value on the quantum
state $\rho$ and $\sigma^2_X$ is the variance of the observable $X$.
The above quantity cannot be applied to eigenstates of $A$ or $B$.
Clearly, ${\cal C}_{AB}=0$ for uncorrelated (product) states.  In
contrast to the classical Pearson correlation coefficient, the quantum
one is, in general, complex if $A$ and $B$ do not commute, but as in
the classical case, its modulus is upper-bounded by one:
\begin{eqnarray}
&&|\<AB\>-\<A\>\<B\>|^2=|\tfrac{\<[A,B]\>+\<\{A,B\}\>}2-\<A\>\<B\>|^2
=\nonumber\\&&|\tfrac12{\<[A,B]\>}|^2+
|\tfrac12{\<\{A,B\}\>}-\<A\>\<B\>|^2\leqslant
\sigma_A^2\sigma_B^2
\labell{pearsbou}\;,
\end{eqnarray}
where $[\cdot,\cdot]$ and $\{\cdot,\cdot\}$ denote the commutator and
anticommutator respectively, and where the final inequality is the 
Schr\"odinger uncertainty relation \cite{schroedinger}. 

We now use $|{\cal C}_{AB}|+|{\cal C}_{CD}|$ as a measure of
complementary correlations to recover some entanglement properties of
the system state. The Pearson coefficient gauges only the linear
correlation of two stochastic variables, so it will not detect maximal
correlation even for a maximally entangled state unless pairs of
observables are linear in each others eigenvalues (e.g.~it would fail
if $A=\sum_jj|a_j\>\<a_j|$ and $B=A^3$). However, if one restricts to
linear observables, one can prove that a state is maximally entangled
if and only if there exist two complementary bases such that $|{\cal
  C}_{AB}|+|{\cal C}_{CD}|=2$, e.g.~if one uses
$A=B=\sum_jj|a_j\>\<a_j|$, $C=D=\sum_jj|c_j\>\<c_j|$ where $|a_j\>$
and $|c_j\>$ are two complementary bases. The proof follows from the
properties of the conditional probabilities (used to prove the
analogous statement for the mutual information) and from the fact that
the Pearson coefficient is $\pm 1$ if and only if there is a
functional relation that connects the two stochastic variables
(details in supplemental material).

Instead, for non maximally entangled states we have two conjectures
which are supported by numerical evidence: (i)~If $|{\cal
  C}_{AB}|+|{\cal C}_{CD}|>1$, the two systems are entangled. As for
the mutual information, the inequality is tight since $\rho_{cc}$ is
separable and has $|{\cal C}_{AB}|+|{\cal C}_{CD}|=1$; (ii)~If $|{\cal
  C}_{AB}{\cal C}_{CD}|>1/4$, the two systems are entangled. Also this
inequality is tight: it is attained by the separable state
$\sum_i(|a_ia_i\>\<a_ia_i|+|c_ic_i\>\<c_ic_i|)/2d$, with $|c_i\>$
eigenstates of $C$. As argued above, the conjecture with the product
is weaker than the one with the sum: proving that all separable states
have $|{\cal C}_{AB}|+|{\cal C}_{CD}|\leqslant 1$ implies $|{\cal
  C}_{AB}{\cal C}_{CD}|\leqslant 1/4$.

The proof of these conjectures is complicated by the fact that the
convexity properties of ${\cal C}_{AB}$ are unknown.  Nonetheless,
they are natural conjectures that are easy to verify for large classes
of states (e.g.~see Fig.~\ref{f:example}b). We have also performed
extensive numerical checks by testing them on large sets of random
states generated according to the prescription described in
\cite{zyc}, and verifying that no state with non-positive partial
transpose \cite{pereshor} lies over the conjectured
threshold.\comment{We can add a figure on this.} 

Note that the Pearson correlation only measures linear correlation,
whereas the mutual information measures all types of correlations. So
one could think that the latter is stronger and that these conjectures
are implied by the mutual information results of the previous section.
Surprisingly, this is false since there exist probability
distributions that have maximal Pearson correlation but negligible
mutual information \cite{pre}.  Indeed, consider the family of
entangled two-qubit states
\begin{eqnarray}
|\psi_\epsilon\>=\epsilon|00\>+\sqrt{1-\epsilon^2}|11\>
\labell{ecco}\;,
\end{eqnarray}
with $\epsilon\in[0,1]$. If one uses $A=B=|1\>\<1|$ and
$C=D=|+\>\<+|$, for all $0<\epsilon<1$ such state has $|{\cal
  C}_{AB}|+|{\cal
  C}_{CD}|=1+2\epsilon\sqrt{1-\epsilon^2}>1$(\footnote{In accordance
  to the result for maximally entangled states, the sum is equal to 2
  only for the maximal entangled state
  $|\Phi^+\>=|\psi_{\epsilon=1/\sqrt{2}}\>$.}), but $|\psi_\epsilon\>$
clearly has negligible mutual information for $\epsilon\to 0$. In
other words, the Pearson correlation identifies $|\psi_\epsilon\>$ as
entangled for all $0<\epsilon<1$ (assuming the above conjectures),
whereas the mutual information does not even identify it as
classically correlated {\em at all} for $\epsilon\to 0$.
\togli{However, }Indeed, numerical simulations suggest that Pearson
correlation is more effective at detecting entanglement in random
states than mutual information.

\togli{If, instead of using the four observables $A,B,C,D$, one uses a set of
generators of $SU(d)$, the separability criterion described in this
section is connected with the correlation-tensor criterion described
in \cite{vicen}.}

\sectionp{Sum of conditional probabilities} The third measure of
correlation we consider is the sum of conditional probabilities ${\cal
  S}_{AB}$, defined as
\begin{eqnarray}
{\cal  S}_{AB}\equiv\sum_ip(a_i|b_i)\;, 
\labell{S_AB}
\end{eqnarray}
where $p(a_i|b_i)$ is the probability of outcome $a_i$ on the first
system conditioned on result $b_i$ on the second. [This is a somewhat
limited measure of correlations as the correspondence
$a_i\leftrightarrow b_i$ among results is clearly arbitrary. A more
relevant measure of correlation should also maximize (or minimize)
over the permutations of the measurement outcomes, but for the sake of
simplicity we will avoid it.] In \cite{spengler} a similar
approach was used, but employing joint probabilities in place of
conditional ones.

Gauging complementary correlations with the sum ${\cal S}_{AB}+{\cal
  S}_{CD}$ we can again obtain information about entanglement and
quantum correlations: (i)~Analogously to the case of the mutual
information, the sum is optimized only for maximally entangled states:
a state is maximally entangled if and only if there exist two
complementary bases such that ${\cal S}_{AB}+{\cal S}_{CD}=2d$;
(ii)~As for the Pearson correlation, we have a conjecture for
non-maximally entangled states: if ${\cal S}_{AB}+{\cal S}_{CD}$ has a
value outside the interval $[1,d+1]$, we conjecture that the two
systems are entangled. As in the previous cases, the inequalities are
tight since the upper bound is attained by the separable state
$\rho_{cc}$ and the lower bound by the separable state
$\sum_i|a_ib_{i\oplus 1}\>\<a_ib_{i\oplus 1}|/d$, with $\oplus$ sum
modulo $d$.

Let us analyse the case of separable states. We remind the reader that
classical-quantum (CQ) and quantum-classical (QC) states have the form
$\sum_ip_i|a_i\>\<a_i|\otimes\rho_i$ and
$\sum_ip_i\rho_i\otimes|a_i\>\<a_i|$ respectively, where
$\{|{a_i}\>\}$ is a set of orthogonal states for one subsystem and
$\{\rho_i\}$ is not an orthogonal set of states.  Note that separable
quantum-quantum (QQ) states comprise all separable ones that are not
CC, CQ or QC.

For these states we can prove that: (iii)~if CC states have maximal
correlations on one of two complementary variables, they are
uncorrelated on the other [formally: if $p(a_i|b_i)=1\ \forall i$ then
we must have $p(c_i|d_i)=1/d\ \forall i$, where $a_i,b_i,c_i,d_i$ are
the results of the measurements of $A,B,C,D$ with $A$ complementary to
$C$ and $B$ to $D$]; \togli{[For qubits, we can state this theorem
  equivalently as: if $p(+|+)=p(-|-)\neq 1/2$ then we cannot have
  maximal correlation on the complementary variables:
  $p(0|0)=p(1|1)\neq 1$.]}  (iv)~CQ states cannot have maximal
correlations on any variable [formally: we cannot obtain $p(a_i|b_i)=
1\ \forall i$, even when $p(c_i|d_i)=1/d$]; (v)~QQ states can have
only {\em partial} correlation for each complementary property.  For
example, the separable two-qubit state
$(|00\>\<00|+|11\>\<11|+{|++\>}{\<++|}+|--\>\<--|)/4$ has partial
correlation on both complementary variables, since
$p(0|0)=p(1|1)=p(+|+)=p(-|-)=3/4$.

Given the properties (iii) and (iv), one might suspect that separable
states with non vanishing quantum correlations have always less
complementary correlations, but this is not the case, as
emphasized by (v). Summarizing, CC states can have maximal correlation
only on one property, CQ states cannot have maximal correlation in any
property, and  QQ states can have some correlation on
multiple properties, but you need pure, maximally entangled states to
get maximal correlations on more than one property.

Regarding the result (i), the proof is a direct consequence of simple
properties of conditional probabilities (see supplemental material) as
for the cases seen previously. The difficulty in proving the
conjecture (ii) stems again from a lack of definite concavity
properties of ${\cal S}_{AB}$, but as for the previous conjecture we
have extensively tested it numerically on random states. One may ask
whether the sum over all outcomes in the statement of the conjecture
is necessary.  Indeed it is: the statement that all separable states
satisfy $1/d\leqslant p(a_i|b_i)+p(c_i|d_i)\leqslant 1+1/d$ for some
$i$ is false (where the two bounds $1/d$ and $1+1/d$ give the bounds
$1$ and $d+1$ we used above when the sum over $i$ is performed). A
counterexample is the separable state $(|00\>\<00|+|++\>\<++|)/2$ of
two qubits for which $p(0|0)+p(+|+)=5/3$. If one uses joint
probabilities in place of conditional ones, a sufficient condition for
entanglement can indeed be proven \cite{spengler}.\comment{can we use
  those results to prove ours? Probably not, because they employ the
  linearity (at least to prove that their relation holds for mixed
  states) and conditional probabilities are nonlinear because of the
  Bayes rule. However, I haven't really thought about this too
  carefully!} The results (iii) and (iv) can be proved at the same
time by using simple properties of CC and CQ states when they are
expressed in two complementary bases (see supplemental material),
whereas property (v) is a direct consequence of the example provided
above.

\sectionp{Extension to more complementary observables} Up to now we
have considered the correlations of the measurement outcomes of two
complementary observables. All systems have at least three
complementary observables \cite{mubs}, and it is known that there are $d+1$ for
$d$-dimensional systems if $d$ is a power of a prime \cite{mub,mubs}.
Our results can be immediately extended to an arbitrary number of
complementary observables by calculating the correlations of all the
known complementary observables and considering the sum of the two
largest ones. For example, for mutual information, we can extend the
condition \eqref{a} to conclude that the state is entangled if
\begin{eqnarray}
&&\nonumber  \mbox{max}(I_{AB},I_{CD},I_{EF},\cdots)+\mbox{max}_2(I_{AB},I_{CD},I_{EF},\cdots)\\&&>\log_2d, 
\labell{mub}\;
\end{eqnarray}
where max$_2$ denotes the second largest term and where $A\otimes B$,
$C\otimes D$, $E\otimes F$, etc. are all observables complementary to
each other.  The extensions of all other results and conjectures are
analogous.

Moreover, at least in the case of qubits the bound at point (ii) for
the mutual information and the conjectured bound at point (i) for the
Pearson correlations can be made stronger by adding correlations for
the third complementary observable. This can improve significantly the
efficiency of the present method if used for entanglement detection.
For details and for a comparison with other known entanglement
detection schemes based on measurements of MUBs see the supplemental
material.

\sectionp{Conclusions}
In summary, we have introduced an interpretation of entanglement
based on classical correlations of the measurement outcomes of 
complementary observables. We have studied different types of correlations 
(mutual information $I$, Pearson coefficient $\cal C$, and sum of conditional
probabilities $\cal S$) for complementary observables of two
systems. We have shown how they provide information on the
entanglement and quantum correlations of a bipartite system. We have
derived the following results and presented a few reasonable
conjectures: (i)~we proved necessary and sufficient conditions for
maximal entanglement for $I$, $\cal C$, $\cal S$, (ii)~we proved
sufficient conditions for entanglement based on $I$ and conjectured
sufficient conditions based on $\cal C$ and on $\cal S$; (iii)~when
gauging complementary correlations using $I$, we proved that the
separable states on the boundary with the entangled-states region are
strictly classically correlated, but the same result is false if one
uses $\cal C$ or $\cal S$; moreover we have shown how $\cal S$
provides insight on CC, CQ, QC, and QQ states, showing that
(iv)~without entanglement only classically correlated CC states can
have maximal correlation on one variable (but then they have no
correlation on the complementary one), whereas (v)~separable QQ states
can have only partial correlations on complementary variables.


One can ask if it is possible to give necessary and sufficient
conditions based on correlations for complementary observables. The
naive statement that entangled states {\em always} have larger
correlations than separable states is false, since it is known that
entangled states exist (e.g.~$|\psi_\epsilon\>$ defined above) that
are arbitrarily close to separable pure states \cite{nielsenkempe} and
to the maximally mixed state (in the sense that for any distance
$\epsilon$ one can choose a sufficiently large dimension $d$ such that
an entangled state is within distance $\epsilon$ from the maximally
mixed state \cite{rubin}). These have vanishing correlations for most
measures of correlation. A notable exception, described above, is the
Pearson coefficient that is able to detect the entanglement of
$|\psi_\epsilon\>$ for all $\epsilon>0$ (but it misses other
types of entangled states).\comment{Perhaps we can think of a necessary
  and sufficient condition that says ``among states with purity $x$,
  entangled states have correlation greater than etc.''}

We acknowledge useful feedback from B. Kraus, K. \u Zyczkowski, and an
anonymous Referee.

\appendix\section{Supplemental Material. Proofs}\label{s:proofs}

\subsection{Sufficient condition for entanglement using mutual
  information}

Here we prove that if $I_{AB}+I_{CD}>\log_2 d$, then the state of the
two systems is entangled. This theorem can equivalently be stated as:
if the state is separable then $I_{AB}+I_{CD}\leqslant\log_2 d$.  

The mutual information is
\begin{eqnarray}
I_{AB}\equiv H(A)-H(A|B)\;, 
\labell{I_AB}
\end{eqnarray}
where $H(A)$ is the entropy of the $A$ measurement outcomes and
$H(A|B)$ is the conditional entropy of the $A$ outcomes, which can be
also written as
\begin{eqnarray}
&&H(A|B)=\\&&
-\sum_{a,b}p(a|b)p(b)\log_2
p(a|b)=\sum_{b}p(b)H(A|B=b)
\nonumber\;,
\end{eqnarray}
where 
\begin{eqnarray}
H(A|B=b)=-\sum_{a}p(a|b)\log_2 p(a|b)
\labell{rel-entb}\;
\end{eqnarray}
is the entropy of the probability distribution $p(a|b)$
for fixed $b$. By definition, separable states can be written as
$\rho=\sum_lp_l\rho_l\otimes\sigma_l$. The conditional state $\rho^{(b)}$
when the result $b$ is obtained from a $B$ measurement on the second 
subsystem is
\begin{eqnarray}
\rho^{(b)}=\sum_l\beta^{(b)}_l\rho_l\;,\
\beta^{(b)}_l=p_l\<b|\sigma_l|b\>/\sum_{l'}p_{l'}\<b|\sigma_{l'}|b\>
\labell{rhob}\;.
\end{eqnarray}
In the above expression for $\beta^{(b)}_l$
the term in the denominator is $p(b)$, namely the probability of getting
outcome $b$ when measuring $B$ on the second subsystem. (Note that, in
contrast to entangled states, the components $\rho_l$ of the first
subsystem have not changed, only the spectrum has changed.)
The concavity of the entropy gives
\begin{eqnarray}
&&H(A|B=b)=H(A)_{\rho^{(b)}}\geqslant\sum_l
\beta^{(b)}_l H(A)_{\rho_l}\labell{conc1}\labell{conc}\\\nonumber
&&\Rightarrow
H(A|B)=\sum_bp(b)H(A|B=b)\geqslant\sum_lp_lH(A)_{\rho_l}
\;,
\end{eqnarray}
where $H(X)_{\rho}$ denotes the Shannon entropy of a measurement of
$X$ on the state $\rho$. The same reasoning for $C$ and $D$ yields
\begin{eqnarray}
H(C|D)\geqslant\sum_lp_lH(C)_{\rho_l}\;.
\labell{concCD}
\end{eqnarray}
Now we use Maassen and Uffink's (MU) entropic uncertainty
relation \cite{mu}, which says that for any state $\rho$ we
have $H(A)_\rho+H(C)_\rho\geqslant-2\ln c$ with
$c=\max_{j,k}|\<a_j|c_k\>|$. For complementary observables, $-2\ln
c=\log_2 d$. This means that
\begin{eqnarray}
&&
H(A|B)+H(C|D)\geqslant\sum_lp_l[H(A)_{\rho_l}+H(C)_{\rho_l}]\nonumber\\&&
\geqslant\log_2 d
,\labell{ineqab}
\end{eqnarray}
where the first inequality is due to the concavity of the entropy, the
second is the MU inequality. The above chain of inequalities and the fact that
$H(A)\leqslant\log_2 d$, $H(C)\leqslant\log_2 d$ imply that
\begin{eqnarray}
&&I_{AB}+I_{CD}=H(A)-H(A|B)+H(C)-H(C|D)\nonumber\\ &&
\leqslant\log_2 d\;,
\labell{bmu}
\end{eqnarray}
which concludes the proof.

\subsection{Maximal mutual information for separable states}
Here we prove that the separable states that satisfy
$I_{AB}+I_{CD}=\log_2 d$, are classically-correlated (CC) zero-discord
states.

The proof given in the previous section employs three inequalities:
(a)~the concavity of the entropy, (b)~the MU inequality, and (c)~the
upper bounds for the entropy $H(A)\leqslant\log_2 d$ and
$H(C)\leqslant\log_2 d$. We want now to find the states $\rho$ for
which the above three inequalities are equalities.

Inequality (b) is an equality only if $\rho_l$ is a pure state and it
is an eigenstate of either $A$ or $C$ \cite{bar}.  We will first
assume that all states $\rho_l$ are eigenstates of $A$, namely
$\rho_l=|a_l\>\<a_l|$.  Moreover, the conditions (c) become equalities
iff the state of the first subsystem (obtained by tracing over the
second) is $\openone/d$.  Therefore, in order to saturate the
inequalities (b) and (c), the separable state must be of the form
$\rho=\tfrac 1d\sum_{l=0}^{d-1} |a_l\>\<a_l|\otimes\sigma_l$, where
$|a_l\>$ are eigenstates of $A$.  Since the entropy is strictly
concave, inequality (a) in Eq.~\eqref{conc1} is saturated only if the
states $\rho_l$ for all $l$ have the same entropy
$H(A)_{\rho_l}=H(A)_{\rho^{(b)}}$ for
the outcomes of measurements of $A$. Since the
$\rho_l$s are all eigenstates of $A$, also $\rho^{(b)}$ must be an
eigenstate of $A$, namely $\rho^{(b)}=|a_{l'}\>\<a_{l'}|$ for some
$l'$ (i.e.~$\beta_l^{(b)}=\delta_{ll'}$). From the definition of
$\beta^{(b)}_l$ in Eq. (\ref{rhob}), this corresponds to $\sigma_l$
being an eigenstate of $B$. Therefore, the form of the input state
that fulfills the above requirements is
\begin{eqnarray}
  \rho_{cc}=\frac 1d\sum_{l=0}^{d-1} |a_l\>\<a_l|\otimes |b_l\>\<b_l|
\labell{cc-state}\;.
\end{eqnarray}
For this state, Eq.~(\ref{concCD}) is then automatically satisfied
with equality because we can write it as
\begin{eqnarray}
  \rho_{cc}=
\tfrac1{d^2}  \sum_l\sum_{jk}|c_j\>\<c_k|e^{i[\theta(a_l,c_j)-\theta(a_l,c_k)]}
  \otimes|b_l\>\<b_l|
  \\
  =\tfrac1{d^2}\sum_{j}|c_j\>\<c_j|\otimes\sum_l |b_l\>\<b_l|
+\sum_i\sum_{j\neq k}\cdots
\labell{cc-state2}\;,
\end{eqnarray}
where $\theta$ is some phase factor and where in the second line we
have separated the part diagonal in $c_j$ (which is the only one that
contributes to the conditional probabilities) from the rest. In this
form, it is clear that the probability distribution of the $C$
measurement is uniform and independent of the outcome of the
measurement $D$ on the second subsystem. Therefore the concavity
condition (a), see Eq.~(\ref{concCD}), gives again an equality. So, if
$\rho_l$ are all eigenstates of $A$, the state that satisfies (a),
(b), and (c) with equality is the classically correlated state
$\rho_{cc}$. The same argument {\em mutatis mutandis} applies to the
case in which all $\rho_l$ are eigenstates of $C$.

In order to complete the proof we have to exclude the case where both
eigenstates of $A$ and $C$ appear in the input state $\rho$, namely a
state of the form
\begin{eqnarray}
  p \sum_{l=0}^{d-1}p_l |a_l\>\<a_l|\otimes\sigma_l +
  (1-p)\sum_{l=0}^{d-1}p'_l |c_l\>\<c_l|\otimes\sigma'_l
\labell{nonopt1}\;
\end{eqnarray}
(which, by an appropriate choice of $p_l$ and $p'_l$ may also describe
states such as
$(|a_0\>\<a_0|\otimes\sigma_0+|c_0\>\<c_0|\otimes\sigma'_0)/2$).  By
taking the partial trace over the second subsystem and then
multiplying to the left by $\<a_m|$ and to the right by $|a_m\>$ for
each $m$, one sees that this state cannot have identity as a marginal
to saturate inequality (c) unless all terms in the first sum are
nonzero. Analogously, using $\<c_m|$ and $|c_m\>$ one sees that also
all terms in the second sum are nonzero.  Now, in order to saturate
the concavity (a) of Eq.~(\ref{conc1}), all states $\rho_l$ must have
the same entropy for measurement of $A$ as $\rho^{(b)}$ but, to
saturate the MU inequality (b), it must also be equal to a pure state
(either an eigenstate of $A$ or of $C$). This is impossible for a
state of the form \eqref{nonopt1} unless $p=0$ or 1, in which case we
go back to the case considered above, that leads to the optimal states
$\rho_{cc}$ of (\ref{cc-state}).

\subsection{Necessary and sufficient conditions for maximal entanglement}
Here we prove the results on maximal entanglement. We start from the
mutual information $I$ and the sum of conditional probabilities $\cal
S$. The case of the Pearson coefficient is treated separately below.

Start by proving that if $I_{AB}+I_{CD}=2\log_2d$ or ${\cal
  S}_{AB}+{\cal S}_{CD}=2d$ then the state is maximally entangled (the
converse will be proven below). From the definition of $I_{AB}$ in Eq.
(\ref{I_AB}), it is upper bounded by $\log_2d$ and can achieve this
bound only when the conditional entropy is null. Since the conditional
entropy is defined as
\begin{eqnarray}
H(A|B)\equiv-\sum_{a,b}p(a,b)\log_2 p(a|b)\;, 
\labell{H(A|B)}
\end{eqnarray}
it is null only when the conditional probabilities are $0$ or $1$
which implies the result for ${\cal S}_{AB}+{\cal S}_{CD}$.  We first
prove this for two qubits for simplicity, with the two complementary
properties identified by projectors on the computational basis
$\{|0\>,|1\>\}$ and on the Fourier basis $\{|\pm\>=|0\>\pm|1\>\}$. We
prove that if $p(0|0)=p(1|1)=p(+|+)=p(-|-)=1$, the state of the two
systems is a maximally entangled state $|\Psi^+\>$. The conditional
probability is $p(0|0)=$Tr$[|0\>\<0|\rho_0]$, where $\rho_0$ is the
state conditioned on obtaining $0$ on the second system:
\begin{eqnarray}
\rho_0=\frac{{}_B\<0|\rho|0\>_B}{\mbox{Tr}[_B\<0|\rho|0\>_B]}
\Rightarrow p(0|0)
=\frac{\<00|\rho|00\>}
{\<00|\rho|00\>+\<10|\rho|10\>}
\labell{r}\;,
\end{eqnarray}
where $\rho$ is the initial state of the composite system.
From the above expression we have that 
$p(0|0)=1$ only if $\<10|\rho|10\>=0$. By the same argument for 
$p(1|1)$ we can
conclude that $\<01|\rho|01\>=0$, so that $\rho$ must belong to the
subspace spanned by $\{|00\>,|11\>\}$. Note that
$\<10|\rho|10\>=\<01|\rho|01\>=0$ implies that also
$\<10|\rho|01\>+\<10|\rho|01\>=0$ (see below). This means that \begin{eqnarray}
\<+\!\!+\!\!|\rho|\!\!+\!\!+\>&\!\!=\!\!&\<00|\rho|00\>+
\<11|\rho|11\>+\<00|\rho|11\>+\<11|\rho|00\>\nonumber\\
\<-\!\!+\!\!|\rho|\!\!-\!\!+\>&\!\!=\!\!&\<00|\rho|00\>+
\<11|\rho|11\>-\<00|\rho|11\>-\<11|\rho|00\>\nonumber\\
\Rightarrow
p(+|+)&\!\!=\!\!&\frac{\<00|\rho|00\>+\<11|\rho|11\>+\<00|\rho|11\>+\<11|\rho|00\>}
{2(\<00|\rho|00\>+\<11|\rho|11\>)}\nonumber,
\end{eqnarray}
which is $p(+|+)=1$ if and only if
$\<00|\rho|00\>+\<11|\rho|11\>=\<00|\rho|11\>+\<11|\rho|00\>$.
The above relation holds if and only if $\rho$ is a pure state, and is equal to
$\rho=|\Psi^{+}\>\<\Psi^{+}|$.  

In order to prove this we employ the inequality
$\rho_{nn}+\rho_{mm}\geqslant\rho_{nm}+\rho_{mn}$ which is an equality
only for states that are pure and balanced when restricted to the
$\{|m\>,|n\>\}$ subspace, namely for states that in such subspace are
$|m\>+|n\>$.  This inequality can be easily proved by decomposing
$\rho=\sum_i\lambda_i|i\>\<i|$ and using the same inequality for pure
states: $|\alpha|^2+|\beta|^2\geqslant\alpha^*\beta+\beta^*\alpha$
(where $\alpha=\<n|\psi\>$, $\beta=\<m|\psi\>$), which can be easily
derived noticing that
$0\leqslant|\alpha-\beta|^2=|\alpha|^2+|\beta|^2-\alpha^*\beta-\beta^*\alpha$,
where equality holds if and only if $\alpha=\beta$. 

We will now extend the above proof to couples of systems in arbitrary finite 
dimension $d$. The matrix elements of the global state $\rho$ in the 
computational basis are denoted for brevity as $\<ij|\rho|kl\>=\rho_{ij,kl}$.
The conditions that the conditional probabilities in the computational basis
must be one are given by
\begin{eqnarray}
p(0|0)=\frac{\rho_{00,00}}
{\sum_{i=0}^{d-1}\rho_{i0,i0}}=1
\labell{rd}\;.
\end{eqnarray}
The above condition is satisfied iff ${\sum_{i=1}^{d-1}\rho_{i0,i0}}=0$, which
implies that $\rho_{i0,i0}=0$ for $i\neq 0$.
The same argument holds for the conditional probability $p(j|j)$ and we then 
have\begin{eqnarray}
p(j|j)=\frac{\rho_{jj,jj}}
{\sum_{i=0}^{d-1}\rho_{ij,ij}}=1
\labell{rjd}\;.
\end{eqnarray}
The above condition is satisfied iff ${\sum_{i\neq j}^{d-1}\rho_{ij,ij}}=0$ 
for fixed $j$, which implies that $\rho_{ij,ij}=0$ for $i\neq j$.
As a consequence, the positivity of  $\rho$ requires that  
$\rho_{ij,kl}=0$ for $i\neq j,k\neq l$. The above conditions then imply that
the only nonvanishing elements of $\rho$ are of the form $\rho_{ii,jj}$.

Let us now denote as $\{|\bar j\>,
j=0,d-1\}$ the Fourier transform of the computational basis, namely\begin{eqnarray}
|\bar j\>=\frac{1}{\sqrt{d}}
\sum_{k=0}^{d-1}e^{i2\pi kj/d} |k\>
\labell{QFT}\;.
\end{eqnarray}The matrix elements in the Fourier transformed basis can
be written as\begin{eqnarray}
&&\rho_{\bar 0 \bar 0,\bar 0 \bar 0}=
\frac{1}{d^2}{\sum_{j,k=0}^{d-1}\rho_{jj,kk}}\\\nonumber
&&\rho_{\bar j \bar 0,\bar j \bar 0}=
\frac{1}{d^2}{\sum_{j,k=0}^{d-1}e^{i2\pi(k-j)\bar i/d} \rho_{jj,kk}}
\labell{rjd0}\;.
\end{eqnarray}
The conditional probability $p(\bar 0|\bar 0)$ then takes the form
\begin{eqnarray}
p(\bar 0|\bar 0)=\frac{\sum_{j,k=0}^{d-1}\rho_{jj,kk}}
{\sum_{\bar i=0}^{d-1}\sum_{j,k=0}^{d-1}e^{i2\pi(k-j)\bar i/d} \rho_{jj,kk}}
\labell{cdf}\;.
\end{eqnarray}
By using the identity
\begin{eqnarray}
\sum_{\bar i=0}^{d-1}e^{i2\pi l\bar i/d}=d \delta_{l,kd}
\labell{iden}\;,
\end{eqnarray}
which means that the l.h.s. of Eq. (\ref{iden}) vanishes for all values of 
$l$ that are not multiples
of the dimension $d$, we can write the denominator in Eq. (\ref{cdf}) as
\begin{eqnarray}
\sum_{j,k=0}^{d-1}\rho_{jj,kk}\sum_{\bar i0}^{d-1}e^{i2\pi(k-j)\bar i/d} 
=d \sum_{j=0}^{d-1}\rho_{jj,jj}
\labell{den}\;,
\end{eqnarray}
By imposing that the probability in Eq. (\ref{cdf}) must be one we have that 
\begin{eqnarray}
p(\bar 0|\bar 0)=\frac{\sum_{j,k=0}^{d-1}\rho_{jj,kk}}
{d\sum_{j=0}^{d-1} \rho_{jj,jj}} =1
\labell{cdfinal}\;.
\end{eqnarray}
By the same reasoning as for the qubit case the above condition
implies that $\rho$ is a projector onto the maximally entangled state
$|\Phi^+\>=\frac{1}{\sqrt{d}}\sum_{j=0}^{d-1}|jj\>$.

To finish the proof for $I$ and $\cal S$, we have to prove the
converses: if the state is maximally entangled, then there exist two
complementary bases such that $I_{AB}+I_{CD}=2\log_2d$ or ${\cal
  S}_{AB}+{\cal S}_{CD}=2d$.  Any maximally entangled state $|\Psi\>$
is local-unitarily equivalent to
$|\Phi^+\>\equiv\sum_j|jj\>/\sqrt{d}$.  Namely, $|\Psi\>=U\otimes
U'|\Phi^+\>$ with $U$ and $U'$ unitaries (which transform local bases
one into another).  The same result as for $|\Phi^+\>$ can then be
achieved by starting from $|\psi\>$ and considering as complementary
bases the ones that are obtained by applying $U^{\dagger}\otimes
U'^{\dagger}$ to the computational and the Fourier bases.

Finally, to prove the necessary and sufficient condition for the
Pearson coefficient $\cal C$, we note that we have just shown that a
state has maximal correlation for two complementary bases if and only
if it is local-unitarily equivalent to $|\Phi^+\>$. Namely, in two
complementary bases, the local outcomes match: ${\cal S}_{AB}+{\cal
  S}_{CD}=2d$.  It is known that the Pearson coefficient takes its
extremal values $\pm 1$ iff the two stochastic variables are linearly
dependent and perfectly correlated (i.e.~in one-to-one
correspondence).\comment{See also testpearson.f}

\subsection{Sum of conditional probabilities for CC and CQ states}
Here we prove that CC states can have maximal correlations only on one
of two complementary variables and that CQ states cannot have maximal
correlations on any variable. The first statement is formalized above
as: if $p(a_i|b_i)=1\ \forall i$ then we must have $p(c_i|d_i)=1/d\
\forall i$, where $a_i,b_i,c_i,d_i$ are the results of the
measurements of $A,B,C,D$.  The second statement is formalized as:
even if $p(c_i|d_i)=1/d\ \forall i$ we still cannot obtain
$p(a_i|b_i)= 1\ \forall i$.

We can prove both statements at the same time by observing that CC and
CQ states can be written in the form
$$\rho_d=\sum_ap_a|a\>\<a|\otimes\rho_a,$$ which can have maximal correlation
on some property only for CC states (where the $\rho_a$ are orthogonal
for different $a$). We can then show that the state $\rho_d$ which
has some correlation on $a$ cannot have any correlation on $c$ when
$c$ is a complementary property, namely if $|\<c|a\>|^2=1/d$. Actually,
\begin{eqnarray}
  \rho_d=
  \sum_a\sum_{cc'}\frac{p_a}d|c\>\<c'|e^{i[\theta(a,c)-\theta(a,c')]}
  \otimes\rho_a
  \\
  =\sum_{c}\frac{1}d|c\>\<c|\otimes\sum_ap_a\rho_a+\sum_a\sum_{c\neq c'}\cdots
\labell{lafs}\;,
\end{eqnarray}
where $\theta$ is some phase factor and where, as above, in the second line we
have separated the part diagonal in $c$ (which is the only one that
contributes to the correlations for $c$) from the rest. It is clear
from the form of the state in \eqref{lafs} that such a state does not
have any correlations in $c$: namely that a measurement of $C$ on the
first system gives no information on the second.

\subsection{Correlations on 3 MUBs for qubits}

Following a suggestion by an anonymous referee, we prove here that the
sufficient condition $I_{AB}+I_{CD}>1$ for entanglement of pairs of
qubits can be made stronger by adding a third MUB for each qubit,
which we name $E$ and $F$. Actually, for a separable two-qubit state
the argument given in the first supplemental section until Eq.
(\ref{concCD}) can be applied also for the third pair of bases $EF$.
We can then write

\begin{eqnarray}
&&H(A|B)+H(C|D)+H(E|F)\geqslant \nonumber\\
&&\sum_lp_l[H(A)_{\rho_l}+H(C)_{\rho_l}+H(E)_{\rho_l}]\;.
\labell{ineqACE}
\end{eqnarray}
We can then exploit a generalisation of the MU inequality to the case
of three MUBs $A,C,E$ for qubits provided in \cite{sanchez}, which says that 
for any qubit state $\rho$ we have $H(A)_\rho+H(C)_\rho+H(E)_\rho\geqslant 2$.
We can then conclude that separable states fulfill the condition
\begin{eqnarray}
I_{AB}+I_{CD}+I_{EF}\leq 1\;.
\labell{bmu1}
\end{eqnarray}

The conjectured bound \begin{eqnarray}
C_{AB}+C_{CD}\leq 1
\labell{pear2}\;
\end{eqnarray}
 for separability in terms of 
Pearson correlations can also be extended to three MUBs, leading to
\begin{eqnarray}
C_{AB}+C_{CD}+C_{EF}\leq 1.
\labell{pear3}\;
\end{eqnarray}
The latter has been tested numerically
(see also the next section). In Fig. \ref{f:exampleextended} we  report the
corresponding results for the mutual information and the Pearson correlations
with three MUBs, to be compared with Fig. \ref{f:example}, where only two
MUBs are considered. As we can see, the improvement is striking. In particular 
note that the Pearson correlations are now suitable to detect all Werner states
since they signature the presence of entanglement for $p>1/3$.

\begin{figure}[hbt]
\begin{center}
\epsfxsize=.9\hsize\leavevmode\epsffile{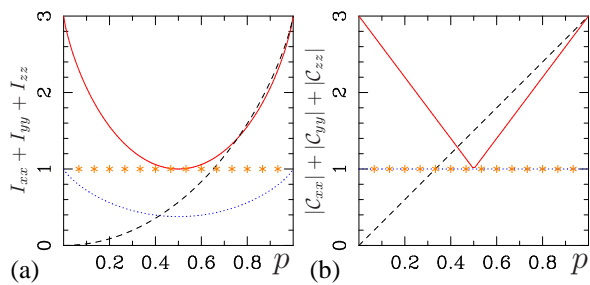}
\end{center}
\vspace{-.5cm}
\caption{
Extension of Fig.~\ref{f:example} to the case in which
  three complementary observables are used for a qubit. (a)~Sum of the
  mutual information $I_{xx}+I_{yy}+I_{zz}$ plotted as a function of
  $p$ for the same families of states presented in
  Fig.~\ref{f:example}.  Note how the criterion now detects a larger
  portion of the Werner states as entangled. The stars indicate the
  threshold above which the states are entangled. (b)~Sum of the
  Pearson correlation coefficients $|C_{xx}|+|C_{yy}|+|C_{zz}|$
  plotted as a function of $p$. The stars indicate the conjectured
  threshold above which states should be entangled. Note that all the
  entangled Werner states (i.e.~the ones for $p>1/3$) are above the
  conjectured threshold: all entangled Werner states are identified by
  the Pearson correlation.  }
\labell{f:exampleextended}\end{figure}

\subsection{Comparison of entanglement detection methods}

In this section we study the effectiveness of our entanglement criterion 
based on complementary correlations to detect entanglement and compare it to
known entanglement detection schemes. We stress that our proposal is
not aimed at introducing a new entanglement detection scheme
but it is aimed at giving a new
interpretation of entanglement in terms of correlations for
complementary properties. Nonetheless, interestingly we see that these
perform better than entanglement witnesses that employ the same
measurements. Moreover, entanglement detection based on correlations
for complementary properties allows to detect many entangled
states that entanglement witnesses miss.

In the following we provide a comparison between our method and an
entanglement detection scheme based on the use of witness operators
\cite{entwitness}. For
simplicity, we compare the case of two qubit states, where
entanglement witnesses that involve measurements of three
complementary properties of qubits take a simple form, namely
\begin{eqnarray}
W_1&=&(\openone+\sigma_x\otimes\sigma_x+
\sigma_y\otimes\sigma_y+\sigma_z\otimes\sigma_z)/4\\
W_2&=&(\openone-\sigma_x\otimes\sigma_x-
\sigma_y\otimes\sigma_y+\sigma_z\otimes\sigma_z)/4\\
W_3&=&(\openone+\sigma_x\otimes\sigma_x-
\sigma_y\otimes\sigma_y+\sigma_z\otimes\sigma_z)/4\\
W_4&=&(\openone+\sigma_x\otimes\sigma_x-
\sigma_y\otimes\sigma_y-\sigma_z\otimes\sigma_z)/4\\
W_5&=&(\openone-\sigma_x\otimes\sigma_x-
\sigma_y\otimes\sigma_y-\sigma_z\otimes\sigma_z)/4
\labell{witness}\;.
\end{eqnarray}
If Tr$[\rho W_i]<0$ for at least one of $i=1,\cdots,5$ for a
two-qubit state $\rho$, then the state is entangled.  
The first four operators are witness operators optimised for the Bell states
\cite{entwitness1}, while $W_5$ was added in order to include also the 
detection scheme  based on the inequality 
$-1\leqslant\<\sigma_x\otimes\sigma_x+
\sigma_y\otimes\sigma_y+\sigma_z\otimes\sigma_z\>\leqslant 1$ for 
separable states \cite{geza}. 
Actually, testing for $W_1$ and $W_5$ allows to test the above condition.
We performed a comparison between these detection schemes and our
criterion (extended to all three complementary bases of qubits as shown in 
the previous section) with a
Monte Carlo simulation, by considering $10^8$ random states obtained
by generating $4\times 4$ random matrices according to the method
detailed in \cite{randomma,randomma1}.  Whether each state is
entangled or not can be evaluated with the PPT criterion
\cite{pereshor}, which is a necessary and sufficient condition in this
case: we found that $36.87\%$ of the states are entangled, which is
consistent with known results \cite{randomma}.  The Pearson
coefficient conjecture detects $9.67\%$ of these entangled states,
whereas the entanglement witnesses detect $8.61\%$ of them. Therefore,
the Pearson coefficient turns out to be more powerful for 
entanglement detection.

In Fig.~\ref{f:venn}a and the first panel of table \ref{ta:pearson} we
compare the performance of these witnesses against the Pearson
correlation conjecture. Note that $24.48\%$ of the detected entangled
states are not detected by the entanglement witnesses and are seen
only by the Pearson coefficient conjecture (this corresponds to a
fraction $28.88\%$ of the entangled stated detected using the Pearson
coefficient: these are detected only by the Pearson correlations). 
The entanglement
witnesses are less effective: they exclusively detect $15.23\%$ of the
detected states (this corresponds to a fraction $20.17\%$ of the
states detected by the entanglement witnesses). In Fig.~\ref{f:venn}b
and the second panel of table \ref{ta:pearson} we compare also the
performance of the sum of conditional probabilities conjecture. The
mutual information does not perform well for entanglement detection,
but interestingly most of the (few) entangled states it detects are
distinct from the ones detected by the witnesses, as seen from the
third panel in table \ref{ta:pearson}.

\begin{figure}[hbt]
\begin{center}
\epsfxsize=.7\hsize\leavevmode\epsffile{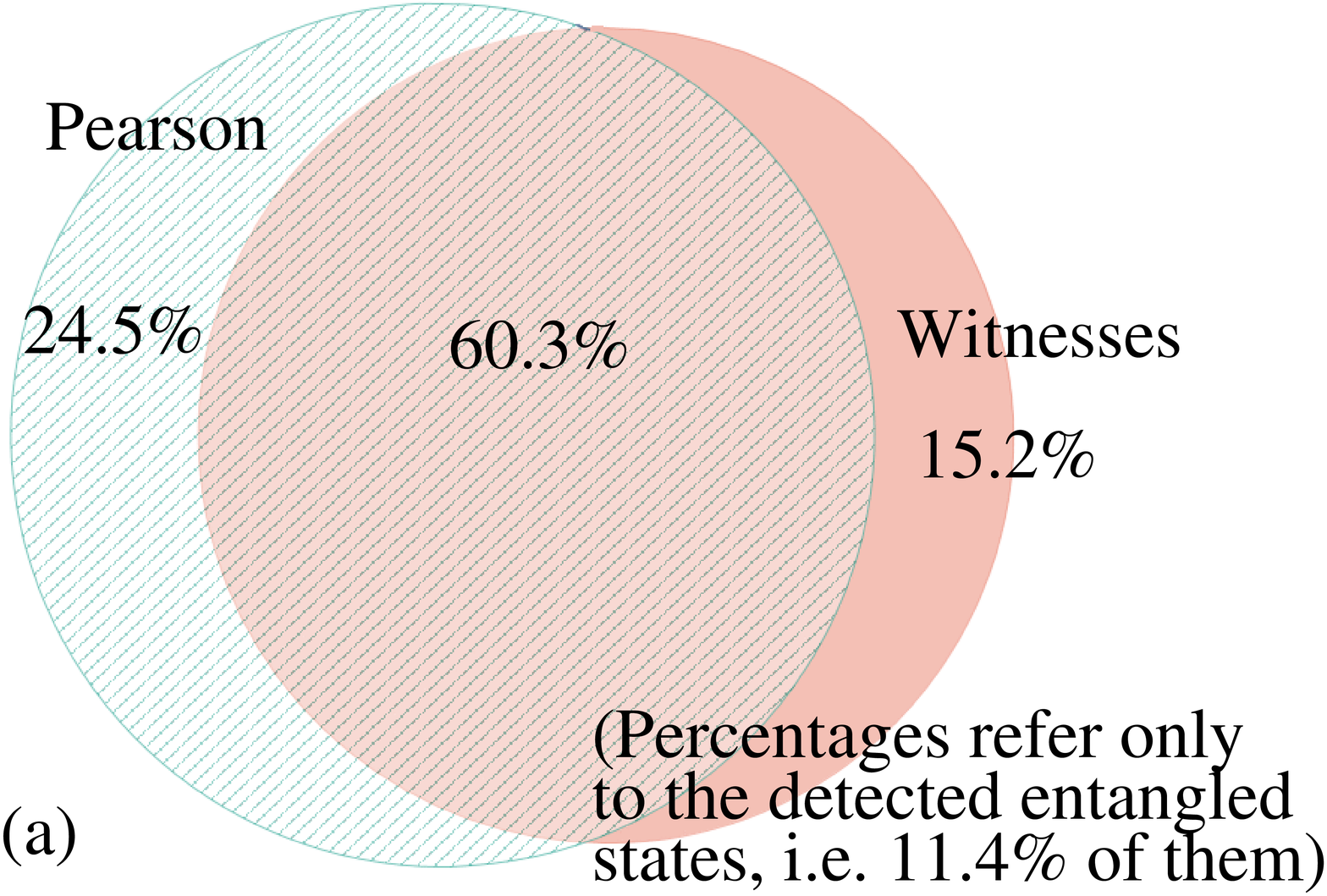}
\epsfxsize=.73\hsize\leavevmode\epsffile{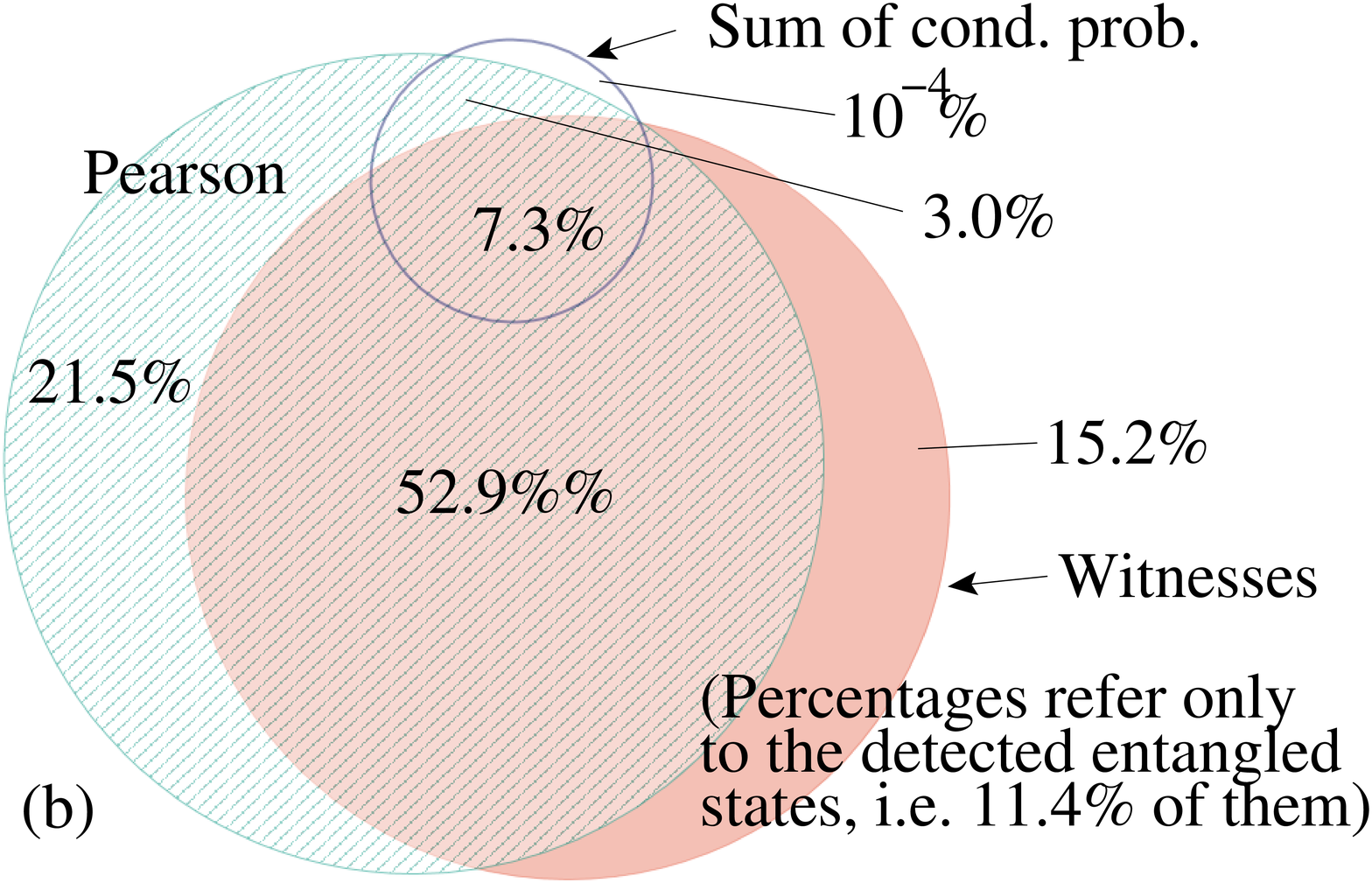}
\end{center}
\vspace{-.5cm}
\caption{Venn diagrams \cite{venn} to compare the efficiency of different
  entanglement detection schemes. (a)~Comparison between the
  entanglement witnesses and the Pearson coefficient conjecture.
  (b)~Comparison between the entanglement witnesses, the Pearson
  coefficient conjecture and the sum of conditional probabilities.}
\labell{f:venn}\end{figure}

\begin{table}\begin{center}\begin{tabular}{|c|c|}\hline
Detection scheme&Entangled states detected\\
\hline\hline
only witness&$15.23\%$\\
only Pearson&$24.48\%$\\
both&$60.29\%$\\
\hline
\hline
 only witness    &$15.23\%$\\
 only Pearson       &$ 21.51\%$\\
 only cond. prob.   &$ 0.0\%$\\
 only witness and Pearson &$52.95\%$\\
only witness and cond.~prob.&$0.0\%$\\
only Pearson and cond.~prob.&$2.98\%$\\
all&   $7.34\%$\\
\hline\hline
 only witness&$15.23\%$\\
 only Pearson&   $24.48\%$\\
 only mutual info& $0.0\%$\\
 only cond.~prob.&$0.0\%$\\
 only witness and Pearson  &$52.95\%$\\
 only witness and mutual info &$0.0\%$\\
 only witness and cond.~prob.&$0.0\%$\\
 only Pearson and mutual info &$0.0\%$\\
only  Pearson and cond.~prob.&$2.78\%$\\
 only mutual info and cond.~prob.           &$0.0\%$\\
 witness, Pearson, mutual info  &$0.001\%$\\
 witness, Pearson, cond.~prob.&$6.72\%$\\
 witness, mutual info, cond.~prob.&$0.0\%$\\
 Pearson, mutual info, cond.~prob.&$0.2\%$\\
 all& $0.61\%$\\\hline
\end{tabular}
\end{center}
\caption{Monte Carlo simulation results. The first panel compares the
  witnesses $W_{1,\cdots, 5}$ with the Pearson conjecture, the second
  panel adds also the 
  sum of conditional probabilities conjecture, the third panel adds
  the mutual information. The first two tables are graphically
  depicted in Fig.~\ref{f:venn}. The percentages depicted refer
  to the fraction of the entangled states that were detected by at
  least one of the methods, which is a fraction of $11.41\%$ of the
  entangled states in all cases since the mutual information and the
  sum of conditional entropies detect very few new entangled
  states that are not detected by the other two schemes: on the order
  of $10^{-4}\%$). We
  sampled  $10^8$  
  random $4\times 4$
  states and
  interpreted them as two-qubit states.
  \labell{ta:pearson}}
\end{table}

We now present a comparison between the Pearson conjecture and a
relation (proposed by an anonymous Referee) based on local uncertainty
relations (LUR) \cite{olga} for qubits. Indeed the LUR implies that
all separable states have
$\Delta^2(\sigma_x\otimes\openone-\openone\otimes\sigma_x)+
\Delta^2(\sigma_z\otimes\openone- \openone\otimes\sigma_z)\geqslant
2$. This is obtained from the uncertainty relation
$\Delta^2(\sigma_x)+\Delta^2(\sigma_z)\geqslant 1$. This condition is
equivalent to saying that all separable states satisfy
\begin{eqnarray}
&&|C'_{XX}|+|C'_{ZZ}|\leqslant\labell{referee}\;
\\&&\nonumber
(\Delta^2_{\rho_1}(\sigma_x)+\Delta^2_{\rho_2}(\sigma_x
)+\Delta^2_{\rho_1}(\sigma_z)+\Delta^2_{\rho_2}(\sigma_z))/2-1,
\end{eqnarray}
where $\rho_1$ is the reduced state over the first subsystem and
$\rho_2$ over the second, and where $C'_{AB}=\langle AB\rangle-\langle
A\rangle\langle B\rangle$. The comparison between this entanglement
detection condition and the Pearson conjecture is presented in
Fig.~\ref{f:referee}, whence it is evident that the Pearson conjecture
\eqref{pear3} that uses three complementary observables identifies
many more entangled states than \eqref{referee}, which can identify
only very few states ($0.39\%$ of the identified ones) that are not
seen by the Pearson conjecture. If one, instead, limits the Pearson
conjecture to the same two complementary observables employed in
\eqref{referee} as in \eqref{pear2}, then the former procedure results
stronger, as it identifies all the entangled states seen.

\begin{figure}[hbt]
\begin{center}
\epsfxsize=.7\hsize\leavevmode\epsffile{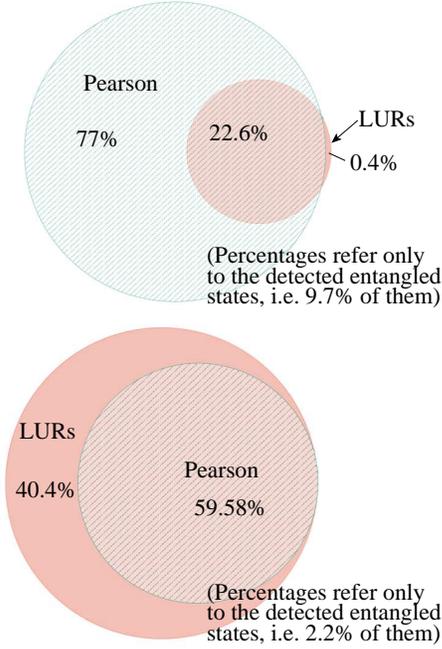}
\end{center}
\vspace{-.5cm}
\caption{Venn diagram \cite{venn} to compare the efficiency of the
  Pearson correlation conjecture and the condition \eqref{referee}
  based on local uncertainty relations (LURs). The data has been
  obtained from a Monte Carlo simulation of $10^8$ random matrices.
  The figure above refers to the Pearson conjecture with three
  complementary observables \eqref{pear3}, the figure below to the
  Pearson conjecture with two \eqref{pear2}.}
\labell{f:referee}\end{figure}

\begin{figure}[hbt]
\begin{center}
\epsfxsize=.9\hsize\leavevmode\epsffile{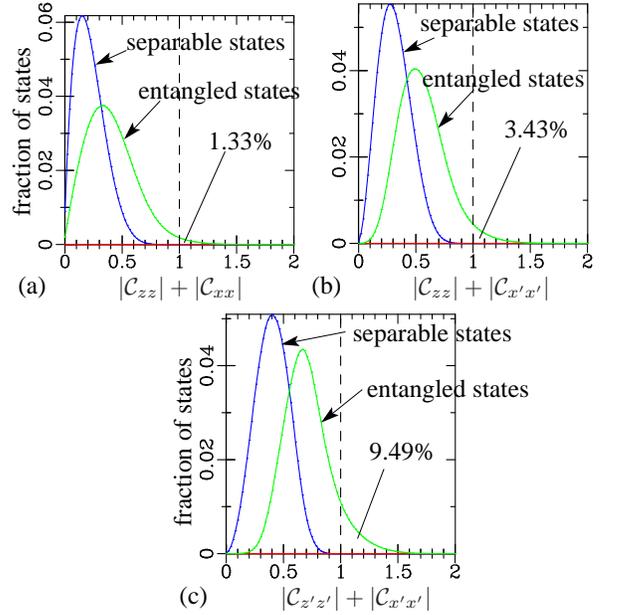}
\end{center}
\vspace{-.5cm}
\caption{Our method seen as an entanglement criterion: effect of the
  optimization over the complementary observables (only two,
  $\sigma_x$ and $\sigma_z$, of the three complementary observables of
  the qubit are optimized here). (a)~Distribution of the Pearson
  correlation $|{\cal C}_{zz}|+|{\cal C}_{xx}|$ for $10^8$ randomly
  generated states without optimization. Only $1.33\%$ of the
  entangled states are above the threshold $|{\cal C}_{zz}|+|{\cal
    C}_{xx}|=1$ (vertical dashed line), and are detected as entangled
  without any optimization. (b)~Same as previous, but optimizing the
  second observable among 80 different directions of the Fourier basis
  (using the same basis for the two qubits). Compared with the
  previous case, note how the optimization reduces the overlap between
  the separable (blue or dark grey) and entangled (green or light
  grey) distributions pushing the entangled distribution over the
  threshold. Now a percentage $3.43\%$ of entangled states are
  detected as such. Here $10^8$ random states have been employed.
  (c)~Double optimization over the two bases (for 40 values of the
  Fourier basis for each of 40 different choices of computational
  basis). The overlap among the histograms is further reduced and now
  a percentage of $9.49\%$ states are detected as entangled. Here
  $10^7$ random states have been employed. A further increase in
  efficiency would be achieved by optimizing over all three
  complementary observables of a qubit, and for different directions
  for each of the two qubits: see Fig.~\ref{f:optim2}.}
\labell{f:optim}\end{figure}

\begin{figure}[hbt]
\begin{center}
\epsfxsize=.9\hsize\leavevmode\epsffile{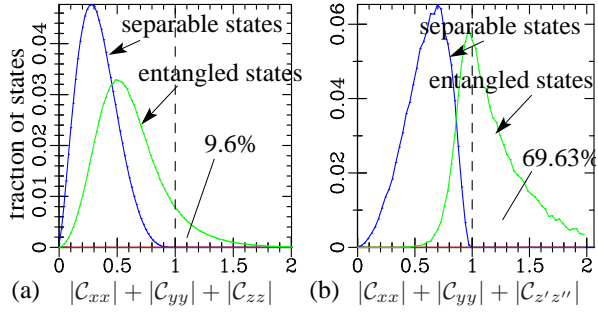}
\end{center}
\vspace{-.5cm}
\caption{Same as Fig.~\ref{f:optim}, but where all three different
  complementary observables of the two qubits have been considered.
  (a)~Distribution of the Pearson correlation $|{\cal C}_{xx}|+|{\cal
    C}_{yy}|+|{\cal C}_{zz}|$ for $10^7$ randomly generated states
  without optimization. Here $9.65\%$ of the entangled states are
  above the threshold $|{\cal C}_{xx}|+|{\cal C}_{yy}|+|{\cal
    C}_{zz}|=1$ (vertical dashed line), and are detected as entangled
  without any optimization, consistently with the results presented in
  the previous section (Fig.~\ref{f:venn}, where
  $84.8\%=60.3\%+24.5\%$ of $11.4\%$ gives indeed $9.66\%$). (b)~Same
  as previous, but optimizing one observable among $10^5$ different
  directions chosen randomly with uniform probability (and
  independently for each qubit). Compared with the previous case, note
  how the optimization reduces the overlap between the separable (blue
  or dark grey) and entangled (green or light grey) distributions
  pushing the entangled distribution over the threshold. Now a
  percentage $69.63\%$ of entangled states are detected as such. Here
  $10^5$ random states have been employed.  \togli{(c)~Double
    optimization over the three complementary observables (for 1000
    values of the first and further 1000 different choices of the
    second, chosen randomly and independently for each qubit). The
    overlap among the histograms should be further reduced but now a
    percentage of $69.04\%$ states are detected as entangled: a
    smaller number of entangled states are detected with respect to
    (b) because numerically it was possible to optimize only over
    fewer directions per qubit: clearly at least as much entangled
    states must be detected here as in the previous case. Here $10^6$
    random states have been employed. (d)~Same as previous for $10^4$
    different complementary observables chosen independently for each
    qubit and $10^4$ random states (the statistical fluctuations arise
    from the smaller number of states employed here).}}
\labell{f:optim2}\end{figure}

\subsection{Efficiency of complementary correlations as  entanglement
  criterion} 
We want to discuss here the efficiency of our proposal as an entanglement 
criterion. It is clear from our proof above
that our criterion could be immediately strengthened by considering
just the entropic uncertainty relations \cite{vittorio} instead of
considering the mutual information: in fact, inequality \eqref{ineqab}
is clearly stronger than \eqref{bmu}. A discussion on these grounds goes
beyond our aims:
we do not want to provide a strong entanglement criterion, but a
different way to interpret entanglement. Nonetheless, the question of
how strong is our proposal as entanglement criterion is a relevant one
and we will explore it in this section.

Let us first distinguish between entanglement detection and
entanglement criteria. The first refers to a measurement procedure,
whose aim is to determine whether a given state is entangled or not
using measurements that are independent of the state. The
second refers to a general characterization of entanglement: how well
does some criterion (in our case, the classical correlation among
complementary observables) characterize entanglement?

It turns out that our proposal is quite effective in characterizing
entanglement. For low dimensions, we explore this numerically for
mixed states.  In Fig.~\ref{f:optim} we explore numerically the use of
the Pearson coefficient conjecture as an entanglement criterion by
using only two MUBs and by optimizing the complementary observables
for each entangled state. For qubits, optimizing only over the second
observable already yields an increase in the detected states: one goes
from a fraction $1.33\%$ of detected entangled states up to $3.43\%$.
An optimization over both observables achieves a further increase up
to a fraction $9.49\%$ of entangled states detected.

Note that the first number $1.33\%$ is lower than the one reported in
the previous section, where we reported that the Pearson correlation
detects a much larger fraction of the entangled states, namely
$84.8\%=60.3\%+24.5\%$ of the detected entangled states which are
$11.4\%$ of the total states, namely $9.7\%=.848\times.114$ of the
total states are detected as entangled. The reason that we achieve
only $1.33\%$ here is because we only considered two instead of the
three complementary observables of a qubit: that optimization would
lead to the increase $1.33\%\to 9.7\%$, as shown in
Fig.~\ref{f:optim2}a.

As mentioned above, in these simulations we have only considered two
complementary observables ($\sigma_z$ and $\sigma_x$). Further
increases can be achieved if one optimizes over all three
complementary observables of qubits: see Fig. \ref{f:optim2}. As
regards to the percentage given in Fig.~\ref{f:optim2}b, we have to
note one should be able to further increase it by optimizing also over
all three complementary observables, instead of only one. Preliminary
investigations in this direction are promising but numerically very
demanding. It is difficult to provide an estimate of the statistical
error since the fluctuations due to the Monte Carlo simulations are
non Gaussian in this case.

Finally, we mention that our method becomes almost an optimal
criterion for large dimensions and pure states according to an
argument pointed out by K. Zyczkowski \cite{karol}.  Unfortunately it
is practically impossible to explore numerically this regime.
Zyczkowski's argument \cite{karol} is based on the fact that for
$d\to\infty$ almost all pure entangled states are close to maximally
entangled, since the average entanglement of randomly chosen
$d$-dimensional pure states goes as $\ln d-1/2+O(\ln d/d)$
\cite{karolpaper}. For large $d$, the term $1/2$ becomes irrelevant.
Since our criterion gives a necessary-and-sufficient condition for
maximally entangled states, it will then asymptotically detect almost
all pure entangled states for $d\to\infty$. This is only a heuristic
argument, however: one must account for the fact that the convergence
to maximal entanglement might be so slow that the above argument might
only be meaningful in the limit $d\to\infty$. In this regime our proof
(valid for finite $d$) does not hold, and discontinuities might be
present \cite{ei}.



\begin{references}
\bibitem{mubs}T. Durt, B.-G. Englert, I. Bengtsson, K. \u Zyczkowski,
Int. J. Quantum Information {\bf 8}, 535
  (2010).
\bibitem{review}D. Bru\ss, 
  J. Math. Phys.  {\bf 43}, 4237 (2002); R. Horodecki, P. Horodecki,
  M.  Horodecki, K.  Horodecki, 
  Rev. Mod. Phys. {\bf 81}, 865 (2009).
\bibitem{pereshor}A. Peres, Phys. Rev. Lett.  {\bf 77}, 1413 (1996);
  M.  Horodecki, P. Horodecki, R. Horodecki, Phys. Lett. A {\bf 223},
  1 (1996).
\bibitem{wern}G. Vidal, R.F. Werner, 
  Phys. Rev. A {\bf 65}, 032314 (2002).
\bibitem{olga}H.F.~Hofmann, S.~Takeuchi, 
  Phys.~Rev.~A {\bf 68}, 032103 (2003).
\bibitem{mahler}J. Schlienz and G. Mahler, 
  Phys. Rev. A {\bf 52}, 4396 (1995).
\bibitem{gunnar}C. Kothe, G. Bj\"ork, 
  Phys. Rev. A {\bf 75}, 012336 (2007); I.S. Abascal, G.  Bj\"ork,
  Phys. Rev. A {\bf 75}, 062317 (2007).
\bibitem{guehne06}O. G\"uhne, M. Mechler, G. T\'oth, P. Adam, Phys.
  Rev. A {\bf 74}, 010301(R) (2006).
\bibitem{zhang07}C.-J. Zhang, Y.-S. Zhang, S. Zhang, G.-C. Guo, Phys.
  Rev. A {\bf 76}, 012334 (2007).
\bibitem{vittorio}V. Giovannetti, Phys. Rev. A {\bf 70}, 012102
  (2004).
\bibitem{gl04}O. G\"uhne and M. Lewenstein, Phys. Rev. A {\bf 70}, 022316
  (2004).
\bibitem{yichen}Y.
  Huang, 
  Phys. Rev. A {\bf 82}, 012335 (2010).
\bibitem{coles}P.J. Coles, M. Piani, 
  Phys. Rev. A {\bf 89}, 022112 (2014); P.J.
  Coles, 
  Phys. Rev. A {\bf 85}, 042103 (2012).
\bibitem{ter}B.M. Terhal, Linear Algebra Appl. {\bf 323}, 61 (2000),
  arXiv:quant-ph/9810091; O. G\"uhne, G. T\'oth, Physics Reports {\bf
    474}, 1 (2009).
\bibitem{cir}M. Lewenstein, B. Kraus, J.I. Cirac, P. Horodecki,
  Phys. Rev. A {\bf 62}, 052310 (2000).
\bibitem{geza} G. T\'oth, 
  Phys.  Rev. A {\bf 71}, 010301(R) (2005); M.R. Dowling, A.C.
  Doherty, S.D.  Bartlett, 
  Phys. Rev. A {\bf 70}, 062113 (2004).
\bibitem{spengler}C. Spengler, M. Huber, S. Brierley, T. Adaktylos,
  B.C. Hiesmayr, 
  Phys. Rev. A {\bf 86}, 022311 (2012); B.C. Hiesmayr, W. L\"offler,
  New J. Phys. {\bf 15}, 083036 (2013).
\bibitem{bill} W.K.~Wootters, 
  Phys. Rev. Lett. {\bf 80}, 2245 (1998).
\bibitem{rudolph}O. Rudolph, Quantum Inf. Process. {\bf 4} 219 (2005),
  arXiv:quant-ph/0202121.
\bibitem{guhne}O. G\"uhne, Phys. Rev. Lett. {\bf 92}, 117903 (2004).
\bibitem{vicen} J.I. de Vicente, 
  Quantum Inf. Comput. {\bf 7}, 624 (2007); J.I. de Vicente, M. Huber,
  Phys.  Rev. A {\bf 84}, 062306 (2011).
\bibitem{eisert}O. G\"uhne, P. Hyllus, O. Gittsovich, J. Eisert,
  Phys.  Rev. Lett.  {\bf 99}, 130504 (2007).
\bibitem{eisert1}O. Gittsovich, O. G\"uhne, P. Hyllus, J. Eisert,
  Phys. Rev. A {\bf 78}, 052319 (2008).
\bibitem{zhang} C.-J. Zhang, Y.-S. Zhang, S. Zhang and G.-C. Guo,
  Phys. Rev. A {\bf 77}, 060301(R) (2008).
\bibitem{wu}S. Wu, Z. Ma, Z. Chen, S. Yu, Sci. Rep. {\bf 4}, 4036
  (2014).
\bibitem{james}J. Schneeloch, C.J. Broadbent, S.P. Walborn, E.G.
  Cavalcanti, J.C. Howell, Phys. Rev. A {\bf 87}, 062103 (2013); J.
  Schneeloch, C.J. Broadbent, J.C. Howell Phys. Rev. A {\bf 90},
  062119 (2014).
\bibitem{mu}H. Maassen, J.B.M. Uffink, Phys. Rev. Lett. {\bf 60}, 1103
  (1988). 
\bibitem{schroedinger}E. Schr\"odinger, 
  Sitzungsberichte der Preussischen Akademie der Wissenschaften,
  Physikalisch-Mathematische Klasse 14: {\bf 296}, 303 (1930).
\bibitem{zyc}K. \u Zyczkowski, P. Horodecki, A. Sanpera, M. Lewenstein,
  Phys. Rev. A {\bf 58}, 883 (1998).
\bibitem{pre}D.V. Foster, P.
  Grassberger, 
  Phys. Rev. E {\bf 83}, 010101(R) (2011).
\bibitem{mub} W. K. Wootters, B. D. Fields, Ann. Phys.  {\bf 191},
  363 (1989).
\bibitem{nielsenkempe}M.A. Nielsen, J. Kempe, 
  Phys. Rev. Lett. {\bf 86}, 5184 (2001).
\bibitem{rubin}A.O. Pittenger, M.H. Rubin, Phys. Rev. A {\bf 62}, 032313
  (2000).
\bibitem{bar} A.R. Gonzales, J.A. Vaccaro and S.M. Barnett, 
Phys. Lett. A {\bf 205}, 247  (1995). 
\bibitem{sanchez} J. Sanchez, Phys. Lett. A {\bf 173}, 233  (1993).
\bibitem{entwitness}O. G\"uhne, P. Hyllus, D. Bru\ss, A.~Ekert,
M. Lewenstein, C.~Macchiavello and A. Sanpera,
Phys. Rev. A {\bf 66}, 062305 (2002).
\bibitem{entwitness1}O. G\"uhne, P. Hyllus, D. Bru\ss, A.~Ekert,
M. Lewenstein, C.~Macchiavello and A. Sanpera,
J. Mod. Opt. {\bf 50}, 1079 (2003).
\bibitem{randomma} K.~\u Zyczkowski, P.~Horodecki, A.~Sanpera,
  M.~Lewenstein,
  Phys. Rev. A {\bf 58}, 883 (1998).
\bibitem{randomma1}M.~Pozniak,  K.~\u Zyczkowski, M.~Kus,  J. Phys. A:
  Math. Gen. {\bf 31}, 1059 (1998).
\bibitem{venn} The Venn diagrams are plotted with the app presented in
  L.  Micallef, P. Rodgers PLoS ONE {\bf 9}, e101717 (2014),
  http://www.eulerdiagrams.org/eulerAPE.
\bibitem{karol}K. \u Zyczkowski, Private communication (during a
  memorable hike in the Iranian mountains) (2014).
\bibitem{karolpaper}K. \u Zyczkowski, H.-J. Sommers, J. Phys. A: Math.
  Gen. {\bf 34}, 7111 (2001),
\bibitem{ei}J. Eisert, C. Simon, M.B.  Plenio, J.  Phys. A: Math. Gen.
  {\bf 35}, 3911 (2002).
\end{references}
\end{document}